\documentclass[prl,preprint,floats]{revtex4}
\usepackage{epsfig}

\begin{document}
\newcommand{\etal}{\textit{et al.\ }}

\title{Hole doping MgB$_2$ without chemical substitution}

\author{Peihong Zhang}
\affiliation{Department of Physics, State University of New York at Buffalo, Buffalo, New York 14260}
\author{Susumu Saito}
\affiliation{
Department of Physics, Tokyo Institute of Technology,
Meguro-ku, Tokyo, 152-8551, Japan}
\author{Steven G. Louie}
\author{Marvin L. Cohen}
\affiliation{
Department of Physics, University of California, Berkeley,
California 94720 and
Materials Sciences Division, Lawrence Berkeley Laboratory, Berkeley,
California 94720}

\date{\today}

\begin{abstract}
{Structures for realizing hole-doped MgB$_2$ without appealing to chemical substitutions
are proposed. These structures which consist of alternating MgB$_2$ and graphene layers have
small excess energy compared to bulk graphite and MgB$_2$. 
Density functional theory based first-principles
electronic structure calculations show significant charge transfer from the MgB$_2$ layer to
graphene, resulting in effectively hole-doped MgB$_2$. Substantial enhancement in the
density of states at the Fermi level of the proposed structure is predicted.}
\end{abstract}
\maketitle

%


Since the discovery of the unusually high superconducting
transition temperature ($T_c=39$ K) in MgB$_2$~\cite{supercond}, 
many attempts have been made to improve its $T_c$ by 
chemical substitutions. It is now well understood that electron
doping through substitution of Mg with Al or B with C
fills the boron $p\sigma$ hole and reduces the electron-phonon (e-ph) coupling 
strength~\cite{Takenobu01,Postorino01,Slusky01,Li02,Pena02,Bianconi02,Mickelson02,Putti03,Wilke04,Zhang05}.
Surprisingly, hole doping 
through substituting Mg with Li or B with Be was also found to
suppress the $T_c$~\cite{Zhao01,Li01,Ahn02}. This is rather unexpected since
it would suggest that MgB$_2$ is naturally optimally doped.

In addition to introducing electrons or holes to the system, however,
chemical substitution also gives rise to
other changes such as impurity scattering~\cite{Kortus05,Sologubenko05,Sologubenko06} and pressure effects 
since the lattice of MgB$_{2}$ may shrink upon substitution. Both of these factors
are detrimental to superconductivity in MgB$_2$ and cannot
be decoupled easily from the intrinsic doping effects.
Furthermore, whereas MgB$_{2}$ is readily doped with electrons,
hole-doping MgB$_{2}$ by substitution of Mg with monovalent ions (e.g., Li or Na)
is more difficult\cite{Cava03,Schilling06}, and there have
been relatively fewer reports on hole-doped MgB$_2$. Detailed study~\cite{Wu05} of
the heat of formation suggests that Li or Na subsituted systems are indeed
unstable or metastable at best. From the viewpoint of the electronic density of states (DOS)
at the Fermi level ($E_F$), another important parameter to
raise $T_c$, hole doping should be more interesting than
electron doping since the DOS curve is a sharply decreasing
function at $E_F$ in MgB$_2$ as will be shown later.
Therefore, realizing ideally hole-doped MgB$_2$ systems is of great interest for
testing theoretical predictions~\cite{Medvedeva01,Choi} as well as achieving potentially higher
$T_c$. 

\begin{figure}
\epsfig{file=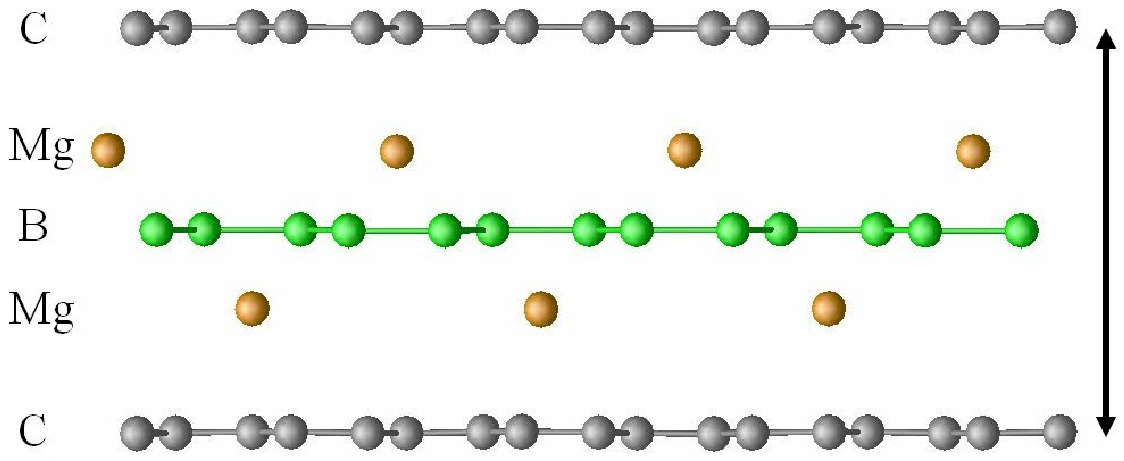, width=7.0cm,angle=0}\\
{(A)}\\
\epsfig{file=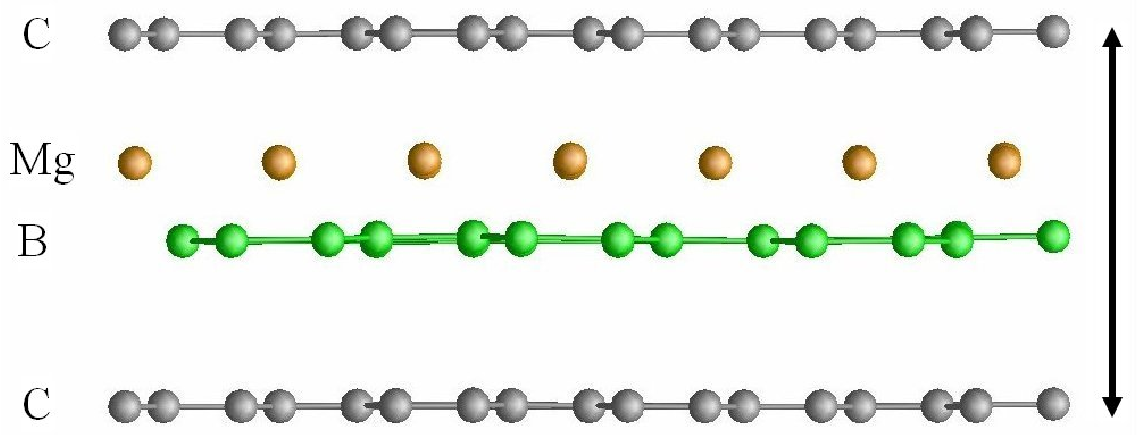, width=7.0cm,angle=0}\\
{(B)}\\
\epsfig{file=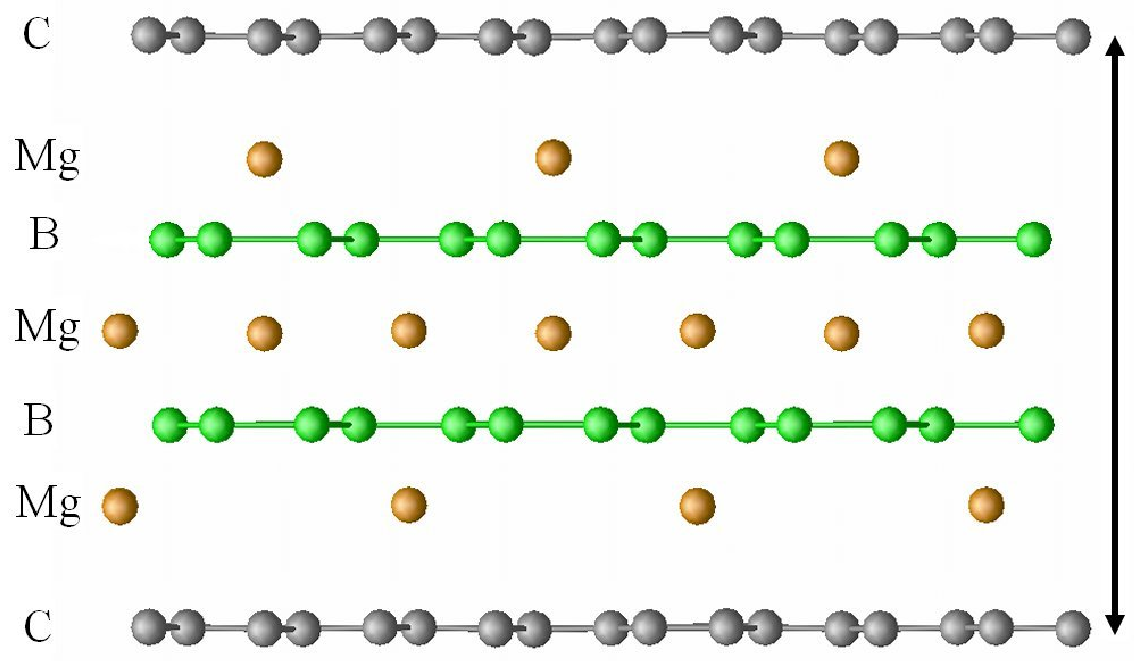, width=7.0cm,angle=0}\\
{(C)}\\
\caption{\label{fig_structure} Proposed composite MgB$_2$/graphite structures. Higher-stage intercalation
structures are also possible.}
\end{figure}

In this paper, we propose novel structures in which MgB$_2$ layers
are {\it effectively} hole-doped without appealing to chemical substitutions.
The proposed structures consist of alternating MgB$_2$ and graphene layers as shown in Fig.\ \ref{fig_structure}. 
Such structures may be realizable using modern deposition techniques. 
For example, a heteroepitaxial double-layer system consisting of a monolayer of boron nitride on Ni(111) and
a graphene overlayer has been reported~\cite{Nagashima96}.
The recent interest in graphite electronics~\cite{Berger06} and the quantum Hall Effect in
graphene~\cite{Novoselov05,YBZhang05} should motivate the development
of novel techniques to control the growth of graphene and related layered systems.


It is well-known that graphite can be doped with electrons by means of metal intercalations. 
Therefore, it is likely that there will be charge transfer from MgB$_2$ to graphite when 
they are stacked together, resulting in effectively hole-doped MgB$_2$. 
We first study two stuctures (Fig.\ \ref{fig_structure}) in which (A) Mg atoms distribute evenly in the B-C interfaces, and 
(B) Mg atoms intercalate into one of the two B-C interfaces. Higher-stage intercalation structures
such as shown in Fig.\ \ref{fig_structure} (C) may also be possible.
All calculations are done using the {\it ab initio} pseudopotential plane wave method~\cite{pseudo} unless
indicated otherwise. The energy cut-off for the plane waves expansion is set at 60 Ry to ensure the
convergence of the calculations since we use norm-conserving pseudopotentials~\cite{PSP2}. The Brillouin
zone is sampled with a uniform $k$-grid with a density equivalent to $24\times 24\times 18$ for a 
primitve MgB$_2$ cell. The calculated in-plane lattice constant $a$ is 2.448 \AA\ for graphite and
3.033 \AA\ for MgB$_2$. It is interesting to note that the ratio
of the in-plane lattice constants $a$(MgB$_2$)/$a$(graphite)$\approx 5/4$. Therefore, we use a supercell
containing $4\times 4$ MgB$_2$ and $5\times 5$ graphene units in our calculations.

All structures studied are fully relaxed within the local density approximation (LDA). 
The average distortion of boron atoms from their ideal positions after relaxation is about $1.2 \times 10^{-3}$ a.u.,
indicating that the underlying hexagonal boron network is minimally perturbed.
The separation between the boron and the carbon layers
in structure A is 3.42 \AA\ whereas those in structure B are 3.05 and 3.79 \AA.
The cohesive energy
of structure A is only 0.25 eV/atom higher compared to those of bulk graphite and MgB$_2$.
Structure B is about 30 meV/atom higher in energy than structure A. 
The stage-2 structure shown in Fig.\ \ref{fig_structure} (C) has an even smaller excess energy of
0.15 eV/atom. These small excess energies indicate the relative stability of the proposed structure. 
With advances in deposition techniques, growing such metastable structues, especially 
higher-stage structures, may be possible. 

\begin{figure}
\epsfig{file=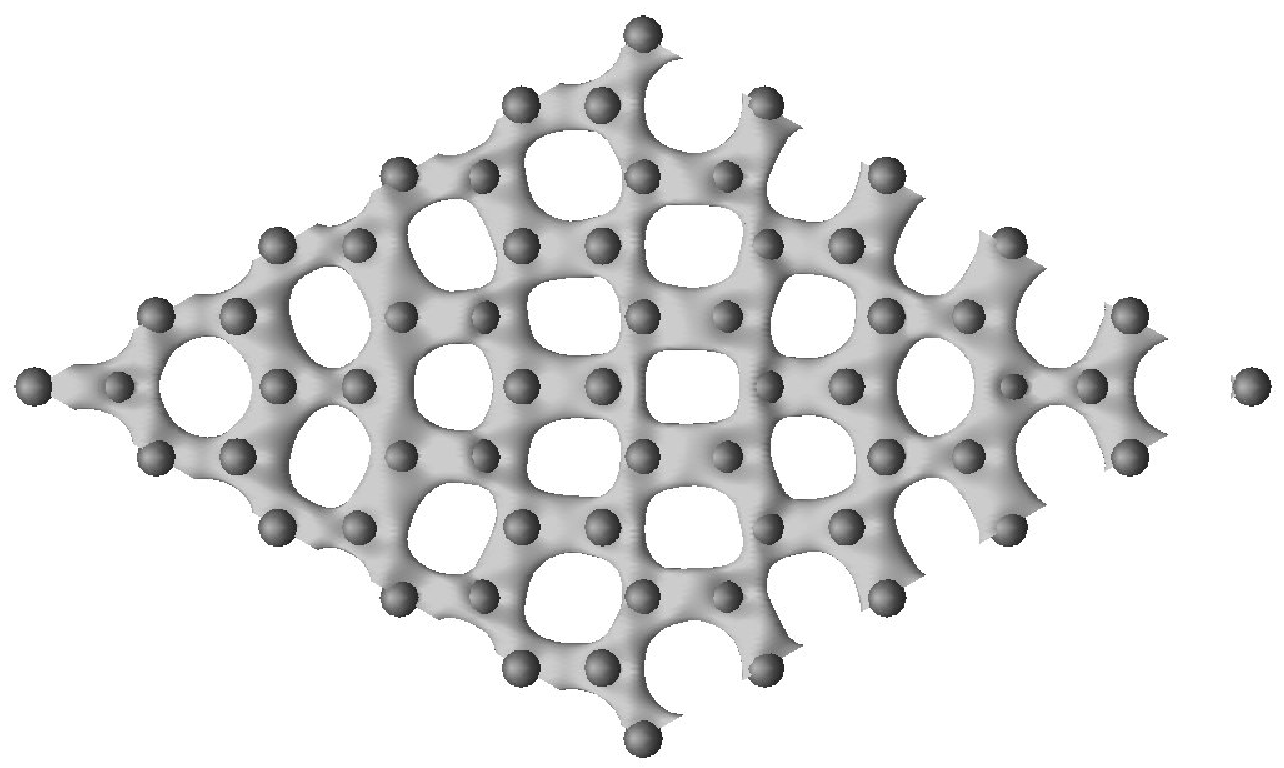, width=7.5cm,angle=0}\\
\caption{\label{fig_charge} Charge density enhancement of the proposed structure B. The 
extra charges are mostly localized near the graphene plane.}
\end{figure}

One of the important results of our study is that the in-plane lattice constants of the proposed structures
expand substantially, 0.9\% for structure A and 1.1\% for structure B. 
It is now well-understood that the C-C bonds in metal intercalated graphite elongate monotonically with increasing 
charge transfer~\cite{Pietronero81,Chan87}. Therefore, the expanded in-plane lattice constant of the alternating
MgB$_2$/C structures suggests that there is significant charge transfer from MgB$_2$ to graphite. To verify this
observation, we have also carried out a calculation using local basis-set as implemented
in the Siesta code~\cite{Siesta}. A standard Mulliken charge analysis shows that the
charge (electron) transfer from MgB$_2$ to graphite is about 0.037 e/C-atom or 0.125 e/MgB$_2$.
The charge transfer effect can also be visualized by plotting the charge density difference
between the proposed composite structures (cs) and the isolated graphite (g) and MgB$_2$ (mb) layers
defined by $\Delta\rho(\mathbf{r})= \rho^{\rm{cs}}(\mathbf{r})-[\rho^{\rm{g}}(\mathbf{r})+\rho^{\rm{mb}}(\mathbf{r})]$.
Figure~\ref{fig_charge} shows the isosurface plot for $\Delta\rho(\mathbf{r})=2.5\times 10^{-3}$ e/a.$\rm{u.}^3$.
It is clear that the charge density enhancement in the composite structures is localized near the carbon plane.


\begin{figure}
\epsfig{file=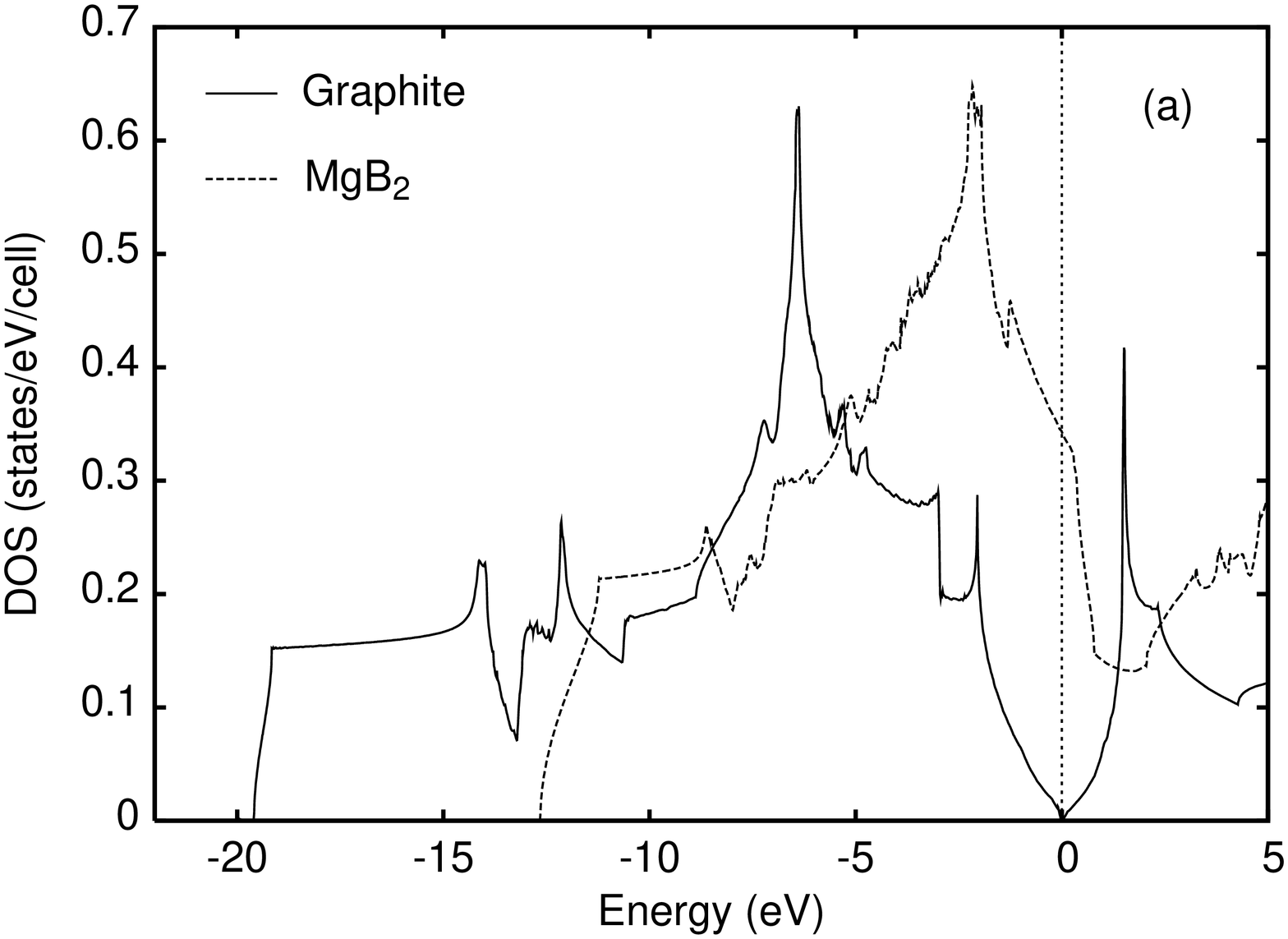, width=6.0cm,angle=-90}\\
\epsfig{file=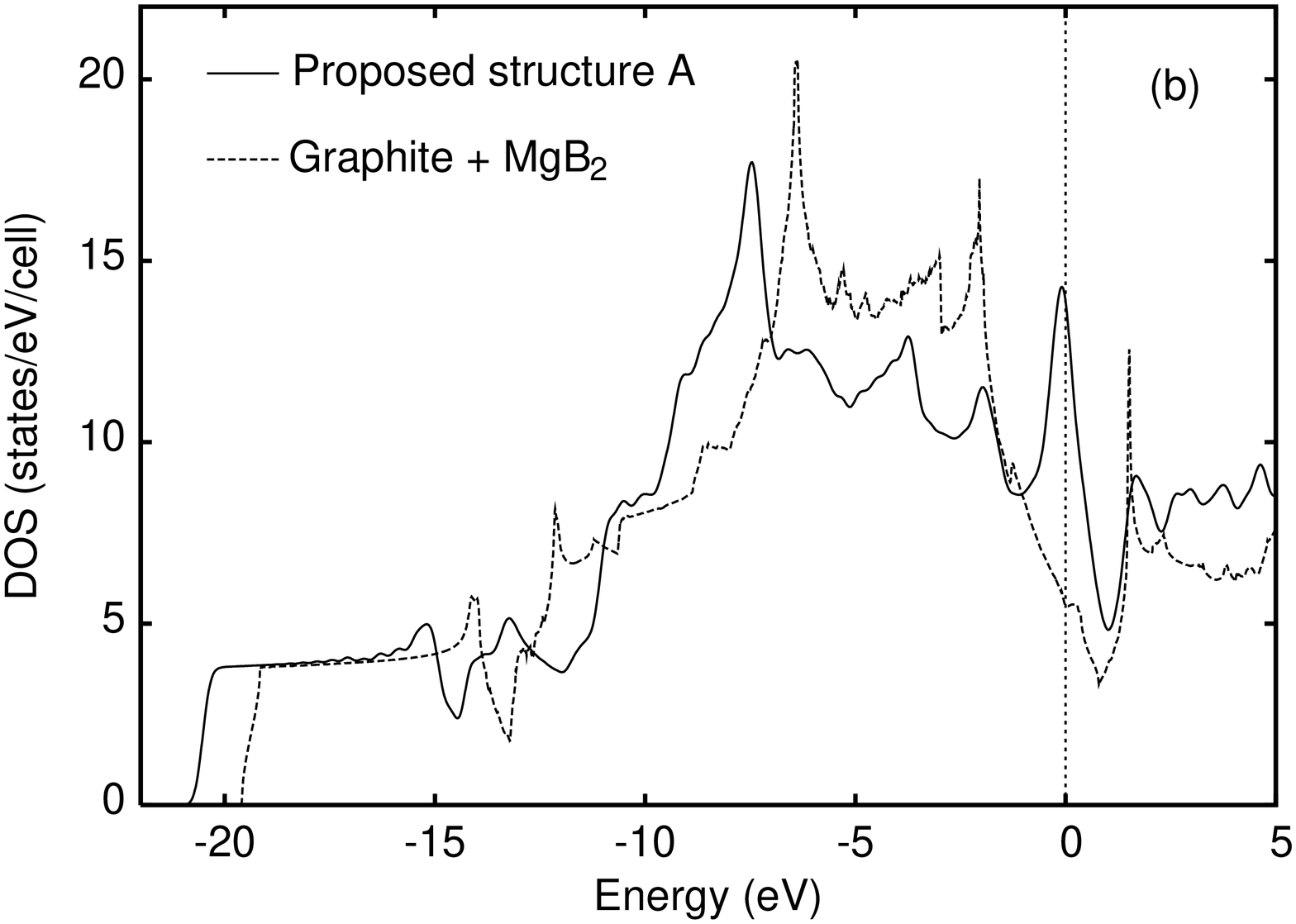, width=6.0cm,angle=-90}\\
\epsfig{file=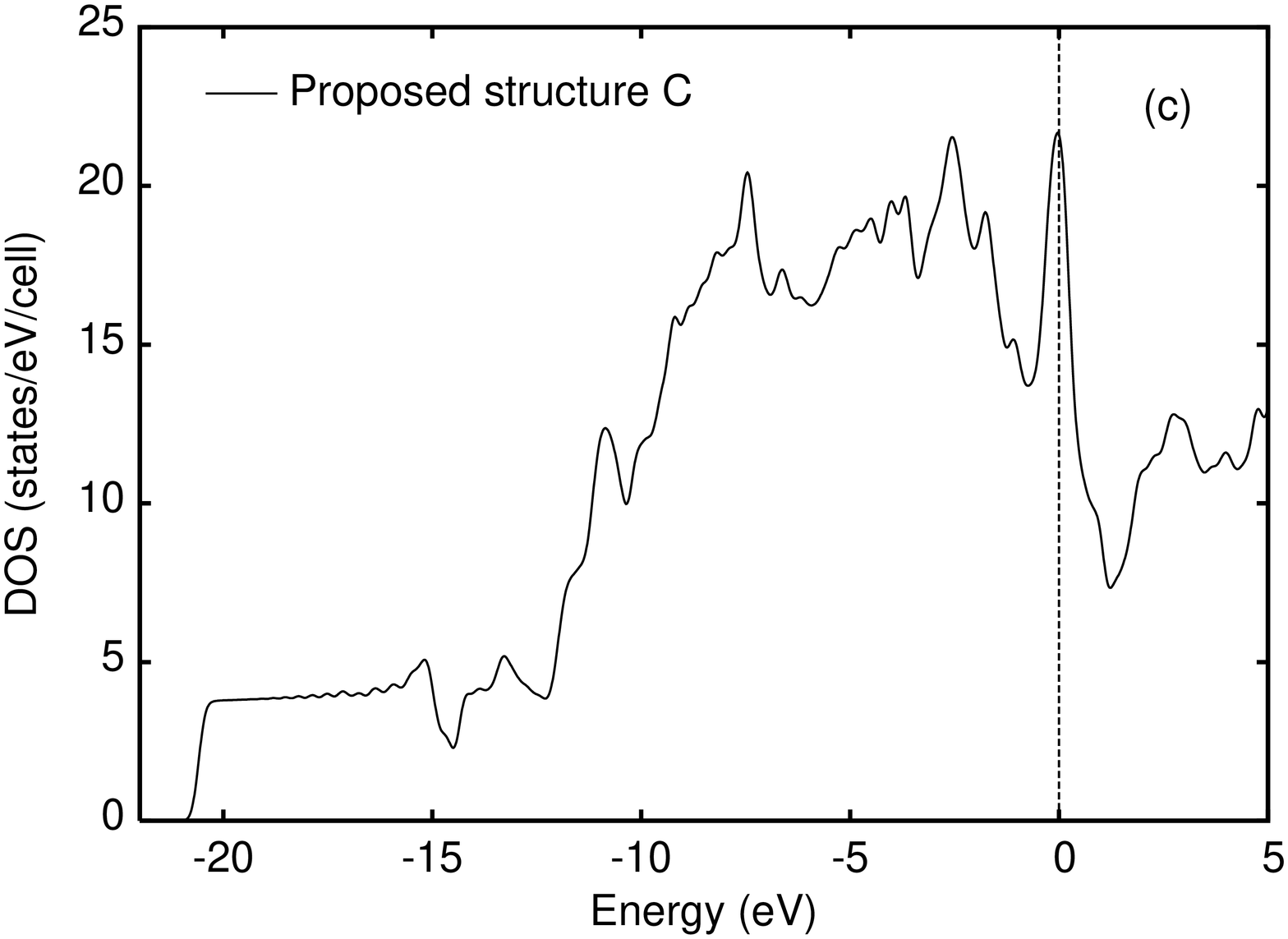, width=6.0cm,angle=-90}\\
\caption{\label{fig_dos} Comparison between the density of states of the proposed structures and those of
bulk graphite and MgB$_2$. The Fermi level $E_F$ is fixed at zero. Significant enhancement in the density of states at the Fermi
level of the proposed structures is largely due to charge transfer effects as explained in the text.}
\end{figure}

To better illustrate the charge transfer effect on the electronic structure,
we compare the DOS for bulk graphite and MgB$_2$ [Fig.\ \ref{fig_dos} (a)],
and that of proposed structure A [Fig.\ \ref{fig_dos} (b)]. The DOS of structure B is similar to
that of structure A; therefore it is not shown. Note that the
DOS of graphite is rescaled to that of a 2-atom cell. Since the DOS of graphite rises sharply
away from the $E_F$, within the rigid-band approximation, we expect a significant 
increase in the DOS at $E_F$ if electrons are transferred to the graphene layer. Similarly, hole-doping
MgB$_2$ will result in an increase in the DOS at $E_F$. 
Indeed, the  calculated DOS at $E_F$ of the proposed structure A is greatly enhanced
compared to the simple addition of the DOSs of the bulk graphite and
MgB$_2$ (without shifting the Fermi levels), as shown in Fig.\ \ref{fig_dos} (b).
In addition, there is a large down-shift ($\sim$ 1.15 eV) of the bottom of the valence band with respect to
the Fermi level in the alternating MgB$_2$/C structure. Since the bottom of
the valence band is primarily of carbon $2s$ character, the down-shift
of these states with respect to the Fermi level is a result of occupying
originally empty $p\pi^*$ states. If we simply integrate the DOS of graphite from
its Fermi level $E_F$ to $E_F + 1.15 eV$, we obtain a charge transfer of 0.039 e/C-atom in
the rigid band approximation. This is in excellent agreement with the Mulliken
charge analysis. Higher-stage structures have similar enhancement in the DOS at
the Fermi level. For example, figure~\ref{fig_dos} (c) shows the DOS 
of the proposed structure C [Fig.\ \ref{fig_structure} (C)] and such an enhancement is evident.

It is widely expected that the e-ph coupling strength should increase
moderately in the ideally hole-doped MgB$_2$, at least for low
doping levels. However, a detailed theoretical understanding of the hole-doped system
is still lacking. On the experimental side, although there have
been reports~\cite{Zhao01,Li01,Ahn02} that the $T_C$ drops in the hole-doped systems, it is
too early to conclude that MgB$_2$ is naturally optimally doped. Difficulties in
synthesizing high quality hole-doped MgB$_2$ through chemical subsitution as well
as impurity scattering effects may offset the intrinsic doping effects. In fact,
even the published results~\cite{Zhao01,Li01} can not be easily reproduced~\cite{Schilling06}. 
Therefore, the proposed structures, if successfully synthesized, are ideal for
studying the ultimate e-ph coupling strenghth in MgB$_2$ and related systems. The significant
enhancement in the DOS of these systems is beneficial for a
stronger e-ph coupling and potentially for a higher $T_c$.

In summary, structures for realizing hole-doped MgB$_2$ without chemical
substitution are proposed. These structures consist of alternating MgB$_2$ and graphene layers. 
Detailed DFT-based electronic structure calculations
indicate that there is a sizeable charge transfer from the MgB$_2$ to graphene 
layeres. In addition, significant enhancement in the DOS of these structures
suggests a stronger e-ph coupling beyond that of ideally hole-doped MgB$_2$.






\begin{acknowledgments}
PZ acknowledges the computational support provided by the Center for Computational
Research at the University at Buffalo, SUNY, and thanks
J. S. Schilling and H. J. Choi for useful discussion.
SS acknowledges the financial support from the Asahi Glass Foundation 
and the 21st Century Center of Excellence Program by Ministry of Education, 
Science, and Culture of Japan through the Nanometer-Scale Quantum Physics 
Project of the Tokyo Institute of Technology.
This work was partially supported by National Science Foundation Grant
No.~DMR00-87088 and by the Director, Office of Science, Office of Basic
Energy Sciences, Division of Materials Sciences and Engineering, U.~S.\
Department of Energy under Contract No.~DE-AC03-76SF00098. 
Computational support from NPACI and by NERSC is also acknowledged.
\end{acknowledgments}

\end{document}